\documentclass[envcountsect]{llncs}

\usepackage{amsmath,stmaryrd}
\usepackage{amsfonts,amssymb}

\usepackage{xy}
\xyoption{all}
 \xyoption{line}
\CompileMatrices

\newcommand{\Set}{{\sf Set}}

\newcommand{\Rel}{{\sf Rel}}

\newcommand{\id}{{\rm id}}

\newcommand{\AAA}{{\cal A}}

\newcommand{\CCC}{{\cal C}}

\newcommand{\FFF}{{\cal F}}

\newcommand{\JJJ}{{\cal J}}
\newcommand{\KKK}{{\cal K}}

\newcommand{\MMM}{{\cal M}}

\newcommand{\RRR}{{\cal R}}
\newcommand{\SSS}{{\cal S}}

\renewcommand{\Bbb}{\mathbb}

\newcommand{\ZZz}{{\Bbb Z}}

\newcounter{countroman}
{\begin{list}{{\rm (\roman{countroman})}}
{\usecounter{countroman}}}%
{\end{list}}

\newcounter{countbarabic}
{\begin{list}{{\rm (\arabic{countbarabic})}}
{\usecounter{countbarabic}}}%
{\end{list}}

\newcounter{countalpha}
\newenvironment{anumerate}%
{\begin{list}{(\alph{countalpha})}{\usecounter{countalpha}}}%
{\end{list}}

\newcounter{countalphabf}
{\protect\begin{list}{{\rm (}{\bf \protect\alph{countalphabf}}{\rm
)}}{\protect\usecounter{countalphabf}}}%
{\end{list}}

\newcommand{\bitmz}{\vspace{-.5\baselineskip}
\begin{itemize}}
\newcommand{\eitmz}{\end{itemize}\vspace{-.
25\baselineskip}}

\newcommand{\bdesc}{\vspace{-.5\baselineskip}
\begin{description}}
\newcommand{\edesc}{\end{description}\vspace{-.
25\baselineskip}}

\mathcode`\<="4268 
\mathcode`\>="5269 
\mathchardef\gt="313E 
\mathchardef\lt="313C 
\newsavebox{\barr}
\savebox{\barr}{\hspace*{-9.5pt}\raisebox{1.25pt}{$
\scriptscriptstyle%
|$}\hspace*{4.5pt}} 
\newsavebox{\barrleft}
\savebox{\barrleft}{\hspace*{-8.5pt}\raisebox{1.25pt}{$
\scriptscriptstyle%
|$}\hspace*{10pt}}

 \def\pushright#1{{
    \parfillskip=0pt            
    \widowpenalty=10000         
    \displaywidowpenalty=10000  
    \finalhyphendemerits=0      
    \leavevmode                 
    \unskip                     
    \nobreak                    
    \hfil                       
    \penalty50                  
    \hskip.2em                  
    \null                       
    \hfill                      
    {#1}                        
    \par}}                      

 \def\qed{\pushright{$\square$}\penalty-700 \smallskip}

\newenvironment{prf}[1]{\begin{trivlist} \item[{\bf ~Proof}#1.]}%
{\qed\end{trivlist}}

\newcommand{\be}[1]{\begin{#1}}

\newcommand{\ee}[1]{\end{#1}}
\newcommand{\beq}{\begin{equation}}
\newcommand{\eeq}{\end{equation}}
\newcommand{\ba}[1]{\begin{array}{#1}}
\newcommand{\ea}{\end{array}}
\newcommand{\bea}{\begin{eqnarray}}
\newcommand{\eea}{\end{eqnarray}}
\newcommand{\bear}{\begin{eqnarray*}}
\newcommand{\eear}{\end{eqnarray*}}
\newcommand{\bpr}{\begin{prf}{}}
\newcommand{\epr}{\end{prf}}
\newcommand{\bprf}[1]{\begin{prf}{#1}}
\newcommand{\eprf}{\end{prf}}

\spnewtheorem*{remarkno}{Remark}{\it}{\rmfamily}
\spnewtheorem*{explanationo}{Explanation}{\it}{\rmfamily}
\spnewtheorem*{propositiono}{Proposition}{\bf}{\it}
\spnewtheorem*{corollaryno}{Corollary}{\bf}{\it}
\spnewtheorem*{lemmano}{Lemma}{\bf}{\it}
\spnewtheorem{thrm}[proposition]{Theorem}{\bf}{\it}
\spnewtheorem{corr}[proposition]{Corollary}{\bf}{\it}
\spnewtheorem{lemm}[proposition]{Lemma}{\bf}{\it}
\spnewtheorem{defn}[proposition]{Definition}{\bf}{\it}

\renewcommand{\to}{\longrightarrow}

\newcommand{\tto}[1]{\xrightarrow{#1}}

\newcommand{\Sto}{{\sf Sto}}
\newcommand{\Ens}{{\sf Ens}}

\newcommand{\Keys}{\KKK}
\newcommand{\Msgs}{\MMM}
\newcommand{\Malg}{\AAA}
\newcommand{\msgs}{M}
\newcommand{\msgp}{\mu}
\newcommand{\Cyph}{\CCC}

\newcommand{\Enc}{{\sf E}}

\newcommand{\Dec}{{\sf D}}
\newcommand{\Attp}{{\sf A}}

\newcommand{\pk}[1]{\overline{#1}}

\newcommand{\Prob}{\Pr}

\newcommand{\code}[1]{\llbracket #1 \rrbracket}

\newcommand{\rand}{\xrightarrow{\RRR}}
\newcommand{\pfn}{\rightharpoonup}
\newcommand{\feas}{\xrightarrow{\FFF}}
\newcommand{\rfeas}{\xrightarrow{\RRR\FFF}}
\newcommand{\ppt}{\xrightarrow{PPT}}
\newcommand{\dpt}{\xrightarrow{DPT}}
\newcommand{\restrictsto}{\upharpoonright}

\newcommand{\from}{\shortleftarrow}

\newcommand{\supp}[1]{\varsigma{#1}}

\title{Chasing diagrams in cryptography\\ \hspace{2em} \\
{\it \normalsize To Jim Lambek for his 90th birthday}}
\author{Dusko~Pavlovic\thanks{Recent primary affiliation: University of Hawaii at Manoa. Email:~{\tt dusko@hawaii.edu}}\\
\email{\small {\rm Email:}~dusko.pavlovic@rhul.ac.uk}}
\institute{Royal Holloway, University of London}

\date{}

\begin{document}

\maketitle

\begin{abstract} \noindent 
Cryptography is a theory of secret functions. Category theory is a general theory of functions. Cryptography has reached a stage where its structures often take several pages to define, and its formulas sometimes run from page to page. Category theory has some complicated definitions as well, but one of its specialties is taming the flood of structure. Cryptography seems to be in need of high level methods, whereas category theory always needs concrete applications. So why is there no categorical cryptography? One reason may be that the foundations of modern cryptography are built from probabilistic polynomial-time Turing machines, and category theory does not have a good handle on such things. On the other hand, such foundational problems might be the very reason why cryptographic constructions often resemble low level machine programming. I present some preliminary explorations towards categorical cryptography. It turns out that some of the main security concepts are easily characterized through \emph{diagram chasing}, going back to Lambek's seminal \emph{`Lecture Notes on Rings and Modules'}.
\end{abstract}

\section{Introduction}
\subsection{Idea}
For a long time, mathematics was subdivided into geometry and arithmetic, later algebra. The obvious difference between the two was that the geometric reasoning was supported by pictures and diagrams, whereas the algebraic reasoning relied upon the equations and abstract text. For various reasons, the textual reasoning seemed dominant in XX century mathematics: there were relatively few pictures in the mathematical publications, and even the formal systems for geometry were presented as lists of formulas. But as the algebraic constructions grew more complex, the task to stratify and organize them grew into a mathematical problem on its own. Category theory was  proposed as a solution for this problem. The earliest categorical diagrams expanded the textual reasoning from exact sequences to matrices of exact exact sequences \cite{MacLane:Homology,Freyd:abelian}. The technique of diagram chasing seem to have emerged around the time of Lambek's classic "Lectures on Rings and Modules" \cite{LambekJ:rings}, where it was used not just as a convenient visualization of lists of equations, but also as a geometric view of universal constructions. This unassuming idea then proceeded to form a germ of geometric reasoning in category theory, uncovering the geometric patterns behind abstract logical structures \cite{PavlovicD:mapsII}. Other forms of geometric reasoning emerged in various forms of categorical research \cite[to mention just a few]{KellyGM:clubs,Joyal-Street:geometry,PavlovicD:Qabs12}, providing some of the most abstract algebraic structures with some of the most concrete geometric tools.

The present paper reports about the beginnings of an exploration towards applying categorical diagrams in a young and exciting area of mathematics: modern cryptography. Initiated in the late 1970s \cite{DH} by introducing algorithmic hardness as a tool of security, modern cryptography developed a rich conceptual and technical apparatus in a relatively short period of time. The increasing complexity of its proofs and constructions, usually presented in a textual, "command line" mode, akin to low-level programming, occasionally engendered doubts that its formalisms may sometimes conceal as many errors as they prevent \cite{another,another-two,another-brave}. Would a high level categorical view help?

\subsection{Background}
Modern cryptography is a theory of effectively computable, randomized boolean functions.  A boolean function is a mapping over bitstrings, i.e. in the form $f : 2^M \to 2^N$, where $2 = \{0,1\}$  denotes the set of two elements, and $M,N$ are finite sets. So $2^M$ denotes the set of $M$-tuples of 0 and 1; or equivalently of the subsets of $M$. Which view of $2^M$ is more convenient depends on the application. Formally, the algebraic structure of $2^M$ is induced by the algebraic structure of $2$, which is usually viewed as
\begin{itemize}
\item Boolean algebra $(2,\wedge, \vee, \neg, 0, 1)$
\item Boolean ring $(\ZZz_2, \oplus, \cdot, 0, 1)$
\item submonoid $\{1,-1\}\subseteq (\ZZz_3,\cdot)$
\end{itemize}
A boolean function $f$ is \emph{effectively computable}, or \emph{feasible}, and denoted by $f:2^M\feas 2^N$, when it is implemented by a boolean circuit, a Turing machine with suitable time and space bounds, or in some other model of computation. Computations in general are, of course, generally expressed as effective boolean functions over the representations of mathematical structures by bitstrings, all the way up to the continuum \cite{PavlovicD:CRN1}.

A \emph{randomized} boolean function $g: 2^M  \rand 2^N$ is in fact a boolean function of two arguments, say $g: 2^R \times 2^M \to 2^N$, where the first argument is interpreted as a \emph{random seed}. The output of a randomized function is viewed as a random variable.  The probability that a randomized boolean function $g$, given an input $x$ produces an output $y$ is estimated by counting for how many values of the random seed $\rho$ it takes that value, i.e. 
\bear
\Prob(y\from gx)  & = &  \frac{\#\{\rho\in 2^R\ |\ y=g(\rho,x)\}}{2^R}
\eear
where $\#S$ denotes the number of elements of the set $S$, and $R$ is the length of the random seeds $\rho$.

An \emph{effective, randomized} boolean function $h: 2^M \rfeas 2^N$ is thus an effectively computable boolean function $h: 2^R \times 2^M \feas 2^N$. It is usually realized by a Deterministic Polynomial-time Turing (DPT) machine, i.e. as $h: 2^R \times 2^M \dpt 2^N$. A DPT with two input tapes, one of which is interpreted as providing the random seeds, is called a Probabilistic Polynomial-time Turing (PPT) machine. So for the same function $h$ we would write $h:2^M\ppt 2^N$, leaving the random seeds implicit. This is  what cryptographers talk about in their formal proofs, although they seldom specify any actual PPTs. Building a PPT is tedious work, in fact an abstract form of low level machine programming. For a high level view of cryptographic programming, an abstract theory of feasible functions is needed.

Before we proceed in that direction, let us quickly summarize what cryptographers actually build from effective randomized boolean functions and PPTs.

A \emph{crypto system} is a structure given over three finite sets
\begin{itemize}
\item $\Msgs$ of {\em plaintexts}
\item $\Cyph$ of {\em ciphertexts}
\item $\Keys$ of {\em keys} 
\end{itemize}
plus a set of random seeds, that we leave implicit. They are all given with their bitstring representations. The structure of the crypto-system consists of three feasible functions
\begin{itemize}
\item key generation $\left<k,\pk k\right> : 1 \ppt \Keys\times \Keys$, 
\item encryption $\Enc :  \Keys \times \Msgs \ppt \Cyph$, and
\item decryption $\Dec : \Keys\times \Cyph\dpt \Msgs$,
\end{itemize}
that together provide
\begin{itemize}
\item { unique decryption: } $ \Dec(\overline{k},\Enc(r,k,m))\  = \ m$,
\item and secrecy.
\end{itemize}
This secrecy is in fact what cryptography is all about. Even defining it took a while. 

The earliest formal definition of secrecy is due to Shannon \cite{Shannon:Secrecy}. His idea was to require that the ciphertext discloses nothing about the plaintext. He viewed the attacker as a statistician, who knows the precise frequency distribution of the language $\Msgs$, i.e. knows for every $m\in \Msgs$ the probability $\Prob(m\from \Msgs)$ that random sampling from $\Msgs$ will yield $m$. Shannon's requirement was that knowing the encryption $c = \Enc(r,k,m)$ of $m$ should not make it any easier to guess $m$, i.e. that
\bea\label{Shan}
\Prob\left( m\from \Msgs\ |\  \exists rk.\ c=\Enc(r,k,m) \right) & = &  \Prob\left(m\from \Msgs\right) 
\eea
Shannon wrote this in a different, but equivalent form\footnote{Except that the encryption was not randomized at the time.}, and called it \emph{perfect security}.

When the age of modern cryptography broke out, the concept of secrecy got refined by considering the feasibility of the encryption and decryption operations, and moreover  strengthened by requiring that the attacker is unlikely to guess not only the plaintext $m$, but even a single bit from it. Otherwise, the concept of secrecy would miss the possibility that the plaintext is hard to guess as a whole, but that it may be easy to guess bit by bit. The original formalization of this requirement is due to Goldwasser and Micali \cite{Goldwasser-Micali82,Goldwasser-Micali84} under the name \emph{semantic security}, but it was later somewhat simplified to the form of \emph{chosen plaintext indistinguishability} (IND-CPA), which looks something like this:
\bea\label{CPA}
\Prob\left( b\from\Attp_1(m_0,m_1,c,s)
 \ \big | \ c\from \Enc(k,m_b),\ b\from 2,\ m_0,m_1,s \from \Attp_0 \right)  
 & \sim &  \frac{1}{2}
\eea
The attacker consists of two PPTs, $\Attp_0$ and $\Attp_1$, which communicate through a tape. She tests the crypto system as follows. First $\Attp_0$ chooses and announces two plaintexts $m_0$ and $m_1$. She may also convey to $A_1$ a part of her state, by writing $s$ on their shared tape. Then the crypto system tosses a fair coin $b$, computes the encryption  $c\from \Enc(k,m_b)$ of one of the chosen plaintexts, and gives it to the attacker. The attacker $\Attp_1$ is now supposed to guess which of the two plaintexts was encrypted. The system is secure if knowing $c$ does not give him any advantage in this, i.e. if his chance to guess $b$ is indistinguishable from $\Prob(b\from 2) = \frac{1}{2}$.

The point that I am trying to make is that this is mouthful of a definition. Especially when we are defining secrecy, which is one of the most basic concepts of cryptography. The upshot is that the most basic cryptographic proofs need to show that some crypto system satisfies the above property.

It is, of course, not unheard of that the fundamental concepts tend to be subtle, and require complicated formal definitions. In cryptography, however, this phenomenon seems to be escalating. First of all, the above definition of secrecy as chosen plaintext indistinguishability turns out to be too weak, and too simple. In reality, the attacker can usually access a decryption oracle, which she can consult before she chooses any plaintexts, and also after she receives back the encryption of one of them, but before she attempts to guess which one it is. So the attacker actually consists of four PPTs, $\Attp_0$, $\Attp_1$, $\Attp_2$ and $\Attp_3$, where $\Attp_0$ begins with choosing some ciphertexts, which it submits to the decryption oracle, etc. A reader who is not a cryptographer may enjoy decyphering the interactions between the crypto system and the attacker from the formula below, describing the \emph{chosen ciphertext indistinguishability} (IND-CCA2), due to Rackoff and Simon \cite{Rackoff-Simon91}. The PPTs again share a tape, which they can use to pass each other a part of the state, denoted $s_0, s_1$ etc. 
\begin{multline}
\Prob\Bigg(b\from\Attp_3(c_0,m,m_0,m_1,c, c_1,\widetilde{m}, s_2)
 \ \bigg | \\ \begin{array}{l} 
m=\Dec(\overline{k},c_0),\ c_0, s_0\from\Attp_0,
\\
c\from \Enc(k,m_b),\  b\from 2,\ m_0,m_1, s_1\from \Attp_1\!
(c_0,m, s_0)\\
\widetilde{m}= \Dec(\overline{k},c_1), \ c_1, s_2 \from \Attp_2(c_0,m, m_0,m_1,c^{\neq}, s_1)
\end{array}\Bigg) \ \  \sim \ \ \frac{1}{2}
\end{multline}
This formula is nowadays one of the centerpieces of cryptography. As verbose as it may look, and as prohibitive as its requirements may be\footnote{The attacker may submit, e.g. two very large plaintexts, say video blocks, as $m_0$ and $m_1$. After she receives the encryption $c$ of one of them, she can then flip just one bit of it, and make that into $c_1$, which is submitted back for decryption. Although $c$ and $c_1$ differ in a single bit, the decryption of $c_1$ should not disclose even a single bit of information about $c_0$.}, it came to be a solid and useful concept. The problem is, however, that the story does not end with it, and that the concepts of ever greater complexity and verbosity rapidly proliferate. This makes cryptographic proofs fragile, with some errors surviving extensive examination \cite{ShoupV:OAEP}. The argument that mandatory formal proofs, if they are too complex, may decrease, rather than increase, the reliability of the proven statements, by decreasing the expert scrutiny over the proven statements, while concealing subtle errors, has been raised from within the cryptographic community \cite{another-Boyd,DentA:another,another-two,another,another-brave}. At the same time, the efforts towards the formalization have ostensibly consolidated the field and clarified some of its conceptual foundations \cite{Goldreich:bookr,Katz-Lindell:book}. Maybe we have good reasons and enough insight to start looking for better notations?

\subsection*{Outline of the paper}
Section~\ref{Symbolic} presents a symbolic model of a crypto system, and a very crude symbolic definition of secrecy. These definitions can be stated in any relational calculus, and thus also in the category of relations. Section~\ref{Information theoretic} presents an information theoretic model of a crypto system. The symbolic definition of secrecy refines here to Shannon's familiar definition of perfect security. We formalize it all in the category of sets and stochastic operators between them. And finally, Section~\ref{Computational} introduces a category where the modern cryptographic concepts can be formalized, such as (IND-CPA) and (IND-CCA2). The upshot of this development is to show how the incremental approach, refining the crude abstract concepts, while enriching the categorical structures, motivates the conceptual development and provides technical tools. Section~\ref{Conclusions} invites for further work.

\section{Symbolic cryptography}\label{Symbolic}
In \cite{Dolev-Yao,Dolev-Even-Karp}, Dolev, Yao, Even and Karp describe public key cryptosystems using an algebraic theory --- roughly what mathematicians would call {\em bicyclic semigroups} \cite{Grillet:semigroups}. 

\subsection{Dolev-Yao crypto systems}

\be{defn}
A {\em message algebra} $\Malg$ consists of three operations:
\begin{itemize}
\item encryption $\Enc : \Malg \times \Malg \to \Malg$, 
\item decryption $\Dec : \Malg \times \Malg \to \Malg$, and
\item key pairing $\pk{(-)}: \Malg\to \Malg$,  
\end{itemize}
and one equation:
\bear
\Dec\left(\pk k, \Enc(k,m)\right) & = &  m
\eear
called {\em decryption condition}. By convention, the first arguments of $\Enc$ and $\Dec$ are called {\em keys}, the second arguments {\em messages}. A message that occurs in $\Enc$ is a {\em plaintext}; a message that occurs in $\Dec$ is a {\em ciphertext}.
\ee{defn}

\be{defn}\label{DY-def} A\/ {\em Dolev-Yao crypto system} is given by 
\begin{itemize}
\item  a  message algebra
\item a set $\msgs \subseteq \Malg$ of {\em well-formed} plaintexts;
\item the {\em hiding condition}: 
"knowing $E(k,m)$ does not reveal anything about $m$"
\end{itemize}
\ee{defn}

\paragraph{Remarks.} The above definitions are close in spirit to Dolev and Yao's definitions, but deviate in details from their presentation. First of all, Dolev and Yao do not present  the encryption and decryption operations as binary operations, but as families of unary operations indexed by the keys. More importantly, their results also require the {\em encryption equation}
\bear
\Enc\left(k,\Dec(\pk k, c)\right) & = &  c
\eear 
that should hold for all keys $k$ and all ciphertexts $c$. Nowadays even toy crypto systems do not satisfy this, so we allow that $\Enc(k,-)$ may not be surjective. Restricted to its image, of course, the decryption equation implies the encryption equation; but not generally. Finally, Dolev and Yao do not take $\msgs\subseteq \Malg$ as a part of the structure. Intuitively, if $\Malg$ is the set of character strings in some alphabet, then $\msgs$ can be construed as the set of words meaningful in some language. For a cryptanalyst, being able to distinguish the meaningful words from the meaningless ones is often critical for recognizing a decryption. The set  $\msgs\subseteq \Malg$ is thus a first, very crude step towards the concepts of source redundancy and frequency distribution, which are of course crucial for cryptanalysis. 

The main challenge left behind Dolev and Yao's analysis is that the hiding condition, which is clearly the heart of the matter, is left completely informal. At the first sight, there seem to be many ways to make it precise. We present one in the next section. Its conceptual analogy with the more familiar information theoretic and computational notions of secrecy are clear, but its technical utility seems limited.

\subsection{Algebraic perfect security}\label{AlgPerfSec}
An attacker sees a ciphertext $c$ and wants to know the plaintext $m$, such that $\Enc(k,m) = c$. But since she does not know the key $k$, she can only form the set of possible\footnote{I.e., this is the only thing that she can do in the \emph{possibilistic} world of mere relations. In the \emph{probabilistic} world of stochastic relations, she can of course do more, and that will be discussed in the next section.} plaintexts $m$ that may correspond to $c$
\bea\label{wideD}
c \widetilde D  & = &  \{m\in \msgs\ |\ \exists k.\ \Enc(k,m) = c\}
\eea
One way to formalize the hiding condition is to require that any well-formed message $m$ must be a candidate for a decryption of $c$, and thus lie in $c\widetilde D$.

%


\be{defn}\label{alg-perfect-def}
A Dolev-Yao crypto system $\Malg$ is \/ {\em algebraically perfectly secure} if every ciphertext can be an encryption of any well-formed message, i.e. if for all $c, m\in \Malg$ there is $k \in \Malg$ such that $\Enc(k,m) = c$.
\ee{defn}
Writing the requirement of this definition in redundant form \bea\label{DY-sec-def}
m\in \msgs\ \wedge \ \exists k\in \Malg.\ \Enc(k,m) = c & \iff & m \in\msgs
\eea
shows that this is a \emph{"possibilistic"}\/ version of \eqref{Shan}. The following lemma says that this captures the intended requirement that the set $c\widetilde D$ does not tell anything about $m$.

\be{lemma}
A Dolev-Yao crypto system $\Malg$ is algebraically perfectly secure if and only  if for all $c, m\in \Malg$ and the binary relation $\widetilde D$ from \eqref{wideD} holds
\bea\label{aperf-lemma}
c\widetilde D m & \iff & m \in \msgs
\eea
\ee{lemma}

A convenient framework to work with algebraic security is the category $\Rel$ of sets and binary relations
\bear
| \Rel | & = & |\Set | \\
\Rel (A,B) & = & \{0,1\}^{A\times B}
\eear
with the usual relational composition of $A\tto R B$ and $B\tto S C$
\bear
a(R;S)c & \iff & \exists b\in B.\  aRb \wedge bSc
\eear
and the equality $aIb \iff a=b$ as the identity relation $A\tto I A$. Note that any subset, say $\msgs \subseteq \Malg$, can be viewed as a relation $\msgs \in \{0,1\}^{1\times \Malg}$, where  $1=\{0\}$, and thus as an arrow $1\tto\msgs \Malg$ in $\Rel$ with $0\msgs x \iff x\in \msgs$.

\be{proposition}
A Dolev-Yao crypto system $\Malg$ is algebraically perfectly secure if and only if the following diagram commutes in the category of relations $\Rel$
\[\xymatrix{
\Malg 
\ar[rr]^{\widetilde\Enc^\msgs} \ar[dd]_{!} & & \Malg\times \Malg \ar[dd]^{!\times \Malg}\\ \\
1 \ar[rr]_{\msgs} && \Malg
}
\]
where \begin{itemize}
\item $\Malg \tto {!} 1$ denotes the total relation, i.e. $x!0$ holds for all $x$ and $1=\{0\}$, and 
\item $\widetilde\Enc^\msgs$ is by definition $c\,\, \widetilde \Enc^\msgs (k,m)\   \iff\   m\in \msgs\wedge \Enc(k,m) = c$.
\end{itemize}
\end{proposition}

\section{Information theoretic cryptography}\label{Information theoretic}
Shannon  \cite{Shannon:Secrecy} brought cryptography to the solid ground of information theory, recognizing the fact that an attacker has access not just to the set $\msgs \subseteq \Malg$ of possible plaintexts, but also to their probabilities $\msgp :\Malg \to [0,1]$. And just like we viewed the former one in the form $\msgs \in \{0,1\}^{1\times \Malg}$ as an arrow $1\tto\msgs \Malg$ in $\Rel$, we shall now view the latter, in the form  $\msgp \in [0,1]^{1\times \Malg}$ as an arrow $1\tto \msgp \Malg$ in the category $\Sto$ of stochastic matrices. 

\subsection{Shannon crypto systems}
To begin, Shannon introduced  into analysis {\em mixed\/} crypto systems, in the form $R = pS + (1-p)T$ where $S$ and $T$ can be thought of as two Dolev-Yao crypto systems, and $p\in [0,1]$. The idea is that the system $R$ behaves like $S$ with probability $p$, and like $T$ with probability $1-p$. In summary, Shannon considered message algebras $\Malg$
\begin{anumerate}
\item given with a probability distribution  $\msgp: \Malg\to [0,1]$ that assigns to each plaintext $m$ a {\em frequency}, $\msgp (m)$, and moreover
\item convex closed, in the sense that for any $p\in [0,1]$
\bear
\Enc\left(pk+(1-p)h, m\right) & = & p\Enc(k, m) + (1-p)\Enc(h,m) \\
\Dec\left(pk+(1-p)h, m\right) & = & p\Dec(k, m) + (1-p)\Dec(h,m)
\eear
\end{anumerate}
But (b) makes it convenient to draw the keys from the  convex hull of $\Malg$ 
\bear
\Delta \Malg & = & \Big\{ \kappa:\Malg\to [0,1]\ \big|\  \#\supp \kappa \lt \infty\ \wedge\ \sum_{x\in \supp \kappa} \kappa(x) = 1\Big\}
\eear
where $\supp \kappa = \left\{x\in \Malg\ |\ \kappa(x) \gt 0\right\}$ is the support. As a consequence, the encryption and decryption maps are not functions any more, but stochastic matrices $\Enc^\kappa$ and $\Dec^\kappa$ with the entries
\bear
\Enc^\kappa_{cm}  & = & \Prob_\kappa(c|m)\ =\  \sum_{\substack{x\in \supp \kappa\\\Enc(x,m) =c}} \kappa(x)\\
\Dec^\kappa_{mc}  & = &\Prob_\kappa(m|c) \ = \  \sum_{\substack{x\in \supp \kappa\\\Dec(\pk x,c) =m}} \kappa(x)
\eear
Condition (a) similarly suggests that a plaintext, or the available partial information about it, should also be viewed as a stochastic vector $\mu\in \Delta \Malg$. A crypto system is now an algebra in the category of sets and stochastic operators
\bear
| \Sto | & = & |\Set | \\
\Sto (M,N) & = & \Big\{\Phi \in [0,1]^{M\times N}\ \big|\ \#\supp \Phi \lt \infty\ \wedge\ \sum_{i\in \supp \Phi} \Phi_{ij} = 1\Big\}
\eear
Indeed, the encryption and the decryption operations are now stochastic operators  $\Enc^\kappa, \Dec^\kappa \in \Sto(\Malg, \Malg)$; whereas the mixed plaintexts  are the points $\mu \in \Sto(1,\Malg)$.

\be{defn}\label{Shannon-def} A {\em Shannon crypto system} is given by 
\begin{itemize}
\item a message algebra in the category $\Sto$, i.e.  stochastic operators for
\begin{itemize}
\item encryption $\Enc : \Malg \times \Malg \to \Malg$, 
\item decryption $\Dec : \Malg \times \Malg \to \Malg$, and
\item key pairing $\pk{(-)}: \Malg\to \Malg$,  
\end{itemize}
\item a frequency distribution of the plaintexts  $\msgp: \Malg\to [0,1]$, and
\item the hiding condition.
\end{itemize}
\end{defn}
This time, the formal definition of the hiding condition available, and well known.

\subsection{Perfect security}
Shannon \cite{Shannon:Secrecy} considers an attacker who makes a {\em probabilistic\/} model of the observed crypto system. More precisely, when she observes a cyphtertext $c$, instead of forming the set $c\widetilde D\subseteq M$  of {\em possible} decryptions, like in Sec.~\ref{AlgPerfSec}, she now tries to compute the conditional distribution $\Prob(m|c)\geq \Prob(m)$ of the {\em probable\/} decryptions of $c$. 

But now the ciphertext $c$ is a random variable $\gamma = \Prob(c)$, which can be viewed as an arrow $1\xrightarrow{\Prob(c)} \Malg$ in $\Sto$. An observation of a ciphertext thus provides knowledge about the distribution of $\Prob(c)$. We assume that the attacker knows the distribution $\kappa = \Prob(k)$ of the keys, and the frequency distribution $\msgp = \Prob(m)$.

\be{defn}\label{prob-perfect-def}
A Shannon crypto system is \/ {\em perfectly secure} if the plaintexts are statistically independent on the ciphertexts, i.e. if for all $c,m\in \Malg$ holds
\bea\label{Sh-sec-def}
\Prob\left(m\from \msgp\ |\ \exists k.\ c= \Enc(k,m)\right) & = & \Prob(m\from \msgp)
\eea
where the conditional probability on the left stands for
\bear
\Prob\left(m\from \msgp\ |\ \exists k.\ c= \Enc(k,m)\right)  & = & \sum_{x\in \supp \kappa} \Prob\left(m\from \msgp  |\ c= \Enc(x,m)\right)\cdot \kappa(x)\notag
\eear
\ee{defn}

Aligning definitions~\ref{alg-perfect-def} and \ref{prob-perfect-def}  shows that algebraic perfect security is an algebraic approximation of Shannon's probabilistic perfect security \cite[II.10]{Shannon:Secrecy}. The following proposition shows that the connection extends to the categorical characterizations.

\be{proposition}
A Shannon crypto system $\Malg$ with finite support is perfectly secure if and only if the following diagram commutes in the category of stochastic operators $\Sto$
\[\xymatrix{
\Malg 
\ar[rr]^{\widetilde \Enc} \ar[dd]_{!} & & \Malg\times \Malg \ar[dd]^{!\times \Malg}\\ \\ 1 \ar[rr]_{\msgp} && \Malg
}
\]
where
\begin{itemize}
\item $\Malg \tto {!} 1$ is the row vector of $\frac{1}{\# \supp\Malg}$,
\item $1\tto \msgp\Malg$ is the distribution $\msgp$ viewed as a column vector,
\item $\Malg \times \Malg \tto{!\times \Malg} \Malg$ is the stochastic matrix with the entries
\bear
(!\times \Malg)_{i(jk)} & = & \begin{cases} \frac{1}{\# \supp \Malg } & \mbox{ if } i=k\\
0 & \mbox{ otheriwise}
\end{cases}
\eear
\item $\widetilde\Enc$ is the stochastic matrix with the entries
\bear
\widetilde\Enc_{c(km)} & = & \begin{cases} \kappa(k)\cdot \msgp(m) & \mbox{ if } c=\Enc(k,m)\\
0 & \mbox{ otherwise}
\end{cases}
\eear
\end{itemize}
\end{proposition}

\section{Computational cryptography}\label{Computational}
Modern cryptography arose from the idea to use computational complexity as a tool, and attacker's computational limitations as the persistent assumptions upon which the cryptographer can built the desired security guarantees. To represent modern crypto system, we need to lift the preceding considerations beyond the mere frequency distributions and randomness, captured in the category $\Sto$, to a category suitable to represent randomized feasible computations, graded by a security parameter.

\subsection{Category of effective stochastic ensembles up to indistinguishability}
The category suitable to present cryptographic constructions will be build by incremental refinement of the category of sets and functions, in three steps: we first make functions feasible, then randomize them, and finally capture the security parameter.

\subsubsection{Effective functions} 
Suppose that every set is given with an encoding: e.g., each element is encoded as a bitstring. A function between encoded sets can then be considered feasible if it is realized by a feasible boolean function on the codes.

Let us begin with a crude realization of this idea, just to get a feeling for it. Let $R = (2^\ast)^{2^\ast}$ be the monoid of boolean functions and $F\subseteq R$ a submonoid of functions that we call \emph{feasible}. For concreteness, we could assume that the functions from $F$ are just those realized by some suitable family of boolean circuits or Turing-machines. The category $\Set_F$ of $F$-computable functions is then defined
\bear
|\Set_F| & = &  |\Set/2^\ast |\ =\ \sum_{A\in |\Set|} \left\{ \code{-}_A:A\to 2^\ast \right\}\\
\Set_F(A,B) & = & \left\{ f\in \Set(A,B)\ |\ \exists \varphi\in F\ \forall a\in A.\ \code{f(a)}_B = \varphi\code a_A\right\}
\eear
\[
\xymatrix{
A \ar[r]^-f \ar[d]_-{\code{-}_A} & B \ar[d]^-{\code{-}_B} \\
2^\ast \ar[r]_-\varphi & 2^\ast
}
\]

\subsubsection{Effective substochastic operators}
Now we want to refine the category $\Set_F$ effective functions to a category of \emph{randomized} effective functions. The step is analogous to the step from $\Set$ to $\Sto$. So randomized effective functions will actually be effective stochastic operators. But since feasible functions may not be total, we will actually work with effective \emph{sub}stochastic operators.

The first task is to define the monoid of randomized boolean functions that will operate on the codes. Consider the set of partial functions
\begin{multline*}
\RRR  =  \{\gamma : 2^\ast \times 2^\ast \pfn 2^\ast\ |\ 
 \forall x\forall \rho_1\forall \rho_2.\ \gamma(\rho_1,x)\!\downarrow\ \wedge\ \gamma(\rho_2,y)\!\downarrow\ \wedge\  |x|=|y|\\  \Longrightarrow\ |\rho_1|=|\rho_2| \ \wedge\  |\gamma(\rho_1,x)| = |\gamma(\rho_2,y)| \big\}
\end{multline*}
where $f(x)\downarrow$ asserts that the partial function $f$ is defined at $x$, and $|\xi|$ denotes the length of the bitstring $\xi$. The set $\RRR$ forms a monoid $(\RRR,\circ,\iota)$ where 
\bea\label{comp}
\gamma \circ \beta(\rho_2 :: \rho_1, x) & = &\gamma(\rho_2, \beta(\rho_1,x))
\eea
and  $\iota(<>,x) = x$, where $<>$ denotes the empty string. This monoid was previously used in \cite{PavlovicD:MFPS10}. Let $\FFF\subseteq \RRR$ be a submonoid of functions that we consider feasible. An example are the functions realized by DPT machines. The category $\Sto_\FFF$ of effective substochastic operators is now defined as follows
\bear
|\Sto_\FFF| & = &  |\Set/2^\ast |\ \ =\ \ \sum_{A\in |\Set|} \left\{ \code{-}_A:A\to 2^\ast \right\}\\
\Sto_\FFF(A,B) & = & \left\{ \Phi\in [0,1]^{A\times B}\ |\ \exists \varphi\in \FFF\ \forall a\in A\ \forall b\in B.\right.\\ && \hspace{13em}\left. \ \Phi_{ab} = \Prob\left(\code b_B \from \varphi\code a_A \right)\right\}
\eear

\subsubsection{Ensembles}
In order to capture security parameters, we must expand randomized functions to ensembles. A \emph{feasible ensemble} is a sequence of feasible functions 
\bear
\psi & = & \left\{\psi_\ell : 2^{r(\ell)} \times 2^{s(\ell)} \feas 2^{t(\ell)}\ |\ \ell \in \omega\right\}
\eear
where $\omega = \{0,1,2,\ldots\}$, and such that
\bear
k\lt \ell & \Longrightarrow &  \psi_k = \psi_\ell \! \restrictsto_{(2^{r(k)}\times 2^{s(k)})}\ \ \wedge\ \ r(k)\lt r(\ell)\ \wedge\  s(k)\lt s(\ell) \ \wedge\  t(k)\lt t(\ell)
\eear
Write $\FFF^\omega$ for the set of feasible ensembles. A typical example of an ensamble is the extensional (i.e. input-output) view of a PPT machine, which can consume longer inputs, and then it produces longer outputs. 

The monoid structure on $\FFF^\omega$ is induced by the monoid structure of $\FFF$. The composite $\vartheta \circ \psi$ of 
\bear 
\vartheta  & = &  \left\{\vartheta_k : 2^{u(\ell)} \times 2^{v(\ell)} \feas 2^{w(\ell)}\ |\ \ell \in \omega\right\}\mbox{ and } \\
\psi  & = &  \left\{\psi_\ell : 2^{r(\ell)} \times 2^{s(\ell)} \feas 2^{t(\ell)}\ |\ \ell \in \omega\right\}
\eear
consists of the components
\bear
(\vartheta\circ \psi)_\ell & = & \overline\vartheta_{\overline \ell} \circ \psi_\ell : 2^{\overline u(\ell)} \times 2^{\overline v(\ell)} \feas 2^{t(\ell)}
\eear 
where 
\begin{itemize}
\item $\overline \ell$ is the smallest number such that $w(\overline \ell ) \geq s(\ell)$, 
\item $\overline \vartheta_{\overline \ell} = \vartheta_{\overline\ell}\restrictsto_{2^{s(\ell)}}$,
\item $\overline\vartheta_{\overline \ell} \circ \psi_\ell$ is defined by \eqref{comp}, 
\item $\overline u(\ell) = u(\overline \ell)$ and $\overline v(\ell) = v(\overline\ell)$.
\end{itemize}
The category $\Sto^\omega_\FFF$ of effective substochastic ensembles is now defined as follows
\bear
|\Ens_\FFF| & = &  |\Set/2^\omega |\ =\ \sum_{A\in |\Set|} \left\{ \code{-}_A:A\to 2^\omega \right\}\\
\Ens_\FFF(A,B) & = & \left\{ \Psi\in [0,1]^{\omega\times A\times B}\ |\ \exists \psi\in \FFF^\omega \ \forall \ell\in \omega\ \forall a\in A\ \forall b\in B.\right.\\
&& \hspace{15em}\left. \Psi^\ell_{ab} = \Prob\left(\code b \from \psi_\ell \code a \right)\right\}
\eear
where
\bear
\Prob\big(\code{b}\from \psi_\ell\left(\code{a}\right)\big) & = & \frac{\#\left\{\rho \in 2^{r(\ell)}\ |\ \code{b}_{t(\ell)} = \psi_\ell\left(\rho, \code{a}_{s(\ell)}\right) \right\}}{2^{r(\ell)}}
\eear
In the special case when $\FFF^\omega$ consists of the actions of PPT machines, we get the category $\Ens_{\rm PPT}$, where the morphisms are the extensional views of PPTs. More precisely, a morphism is a sequence of substochastic matrices $\Psi = \{\Psi^\ell\}_{\ell\in \omega}$ such that there is a PPT $\Pi$ and the $ab$-entry of $\Psi^\ell$ is $\Psi_{ab}^\ell = \Prob(b\from \Pi_\ell a)$, where $\ell$ is the security parameter.

So $\Ens_\FFF$ comes close to providing an abstract view of the universe in which the cryptographers work. The view is abstract in the sense that $\FFF$ does not have to be realized by PPTs, but can be any submonoid of $\RRR$. By taking $\FFF$ to be the PPT realized stochastic operations we get the usual probabilistic algorithms --- \emph{except} that those that are indistinguishable, because their difference is a negligible function still correspond to different morphisms in $\Ens_{\rm PPT}$.

\subsubsection{Indistinguishability}
Note, first of all, that $[0,1]$ is not only a monoid, but an ordered semiring\footnote{A \emph{semiring} is a structure $(R,+,\cdot,0,1)$ such that $(R,+,0)$ and $(R,\cdot,1)$ are commutative monoids such that $a(b+c) = ab+ac$ and $a0 = 0$.}. The semiring structure lifts to $[0,1]^\omega$. A \emph{semi-ideal} in an ordered semiring is a lower closed subset closed under addition and multiplication. Since it is lower closed, it contains 0, but generally not 1. 

Let $\Upsilon \subseteq [0,1]^\omega$ be a semi-ideal. The canonical example is the semi-ideal of \emph{negligible functions} \cite{Goldreich:bookr}. A function $\nu:\omega \to [0,1]$ is called negligible if $\nu(x) \lt \frac{1}{q(x)}$ holds eventually, for every positive polynomial $q$. Any semi-ideal $\Upsilon$ induces on $[0,1]^\omega$ the equivalence relation
\bear
\sigma \underset{\Upsilon}\sim \tau & \iff & \exists \nu \in \Upsilon.\ |\sigma_\ell - \tau_\ell |\lt \nu(\ell) 
\eear
and we define
$\Ens_\FFF^\Upsilon$ to be the category with the same objects as $\Ens_\FFF$, but
\bear
\Ens_\FFF^\Upsilon (A,B) & = & \Ens_\FFF(A,B)\big/ \underset{\Upsilon}\sim
\eear
Unfolding this definition over the semring $\JJJ_\Upsilon = [0,1]^\omega / \underset{\Upsilon}\sim$, we have
\bear
\Ens_\FFF^\Upsilon(A,B) & = & \left\{ \Psi\in \JJJ_\Upsilon^{A\times B}\ |\ \exists \psi\in \FFF^\omega \ \forall \ell\in \omega\ \forall a\in A\ \forall b\in B.\right.\\
&& \hspace{13em}\left. \Psi^\ell_{ab} = \Prob\left(\code b \from \psi_\ell \code a \right)\right\}
\eear

\subsection{Characterizing semantic security}
The usual definition of a crypto system from the Introduction can now be stated abstractly, in a categorical form. While the definition follows the pattern of Def.~\ref{DY-def} and Def.~\ref{Shannon-def}, this time we revert to the usual \emph{multi-sorted} specification, where the plaintexts, the ciphertexts and the keys are drawn from different sets.

\be{defn} An {\em abstract crypto system}, relative to a monoid $\FFF$ of feasible functions, and a semi-ideal $\Upsilon$ of negligible functions is given by 
\begin{itemize}
\item a multi-sorted message algebra in the category $\Ens_\FFF^\Upsilon$, such that
\begin{itemize}
\item encryption $\Enc : \Keys \times \Msgs \to \Cyph$, is a stochastic ensemble, whereas
\item decryption $\Dec : \Keys \times \Cyph \to \Msgs$, and
\item key pairing $\pk{(-)}: \Keys\to \Keys$ are deterministic functions\footnote{Deterministic functions can be characterized intrinsically in $\Sto_\FFF$, $\Ens_\FFF$ and $\Ens_\FFF^\Upsilon$.}.
\end{itemize}
\item a frequency distribution of the plaintexts  $\msgp: \Msgs\to [0,1]$, and
\item the hiding condition.
\end{itemize}
\end{defn}
\paragraph{The upshot of it all.} The abstract versions of the hiding conditions, such as (IND-CPA) and (IND-CCA2), described in the Introduction, boil down to commutative diagrams in $\Ens_\FFF^\Upsilon$. We illustrate this fact for (IND-CPA).

\be{proposition}
Let $\Ens_{\rm PPT}^{\displaystyle \nu}$ be the category of ensembles of PPT-realized boolean functions modulo negligible functions. A crypto system in the usual sense (as described in the Introduction) is equivalent to an abstract crypto system in this category. Such a crypto system is semantically secure, i.e. it satisfies (IND-CPA), as defined by \eqref{CPA}, if and only if the following diagram commutes for all arrows $\Attp_0$ and $\Attp_1$ in $\Ens_{\rm PPT}^{\displaystyle \nu}$.
\[\xymatrix@C-1pc{
\Keys  \ar[rr]^-{<\id_\KKK, \Attp_0>} \ar[dd]_{!} & & \Keys\times \Msgs^2 \times \SSS \ar[rrr]^-
{\pi} 
&&& \Keys \times \Msgs \times \Msgs^2 \times \SSS \ar[rr]^-{\Enc\times \id} && \Cyph \times \Msgs^2\times \SSS \ar[dd]^{\Attp_1}\\ \\
1 \ar[rrrrrrr]_{b} && && &&& 2
}
\]
\end{proposition}

A similar proposition holds for (IND-CCA2).

%
%

\section{Further work}\label{Conclusions}
While the various notions of secrecy can thus be characterized by commutative diagrams in suitable categories, the notions of one-way function and pseudo-random generator correspond to the requirements that some diagrams \emph{do not} commute. This leads to interesting categorical structures, which seem to be best expressed in terms of enriched categories, and the suitable convolution operations. This observation led to an different approach, through \emph{monoidal computer}  \cite{PavlovicD:MSCS97,PavlovicD:IC12}, lifting the ideas from another strand of Lambek's work, leading from infinite abacus as an intensional model of computation \cite{LambekJ:abacus},  to the extensional models  \cite{LambekJ:Advances}, elaborated in the book with P.J.~Scott \cite{Lambek-Scott:book}.

But how useful might our categorical models of computation be for cryptography? Can the categorical tools, developed for high level program semantics, really be used to stratify cryptographic constructions? The preliminary evidence, some of which was presented here, suggests that certain types of cryptographic proofs and constructions can be significantly simplified by using categorical tools to `hide the implementation details'. The price to be paid, though, is that this hiding requires some preliminary work. For instance, we have seen that the secrecy conditions can be captured by simple diagrams, albeit in randomized categories. This approach echoes the well established programming methodologies, where complex structures are encapsulate into components that hide the irrelevant implementation details, and only the fragments that need to be manipulated are displayed at the interface. The categorical approach developed in Lambek's work has made such strategies available across a broad gamut of sciences. 
\bibliography{ref-catcry,PavlovicD}

\begin{thebibliography}{10}

\bibitem{another-Boyd}
Kim-Kwang~Raymond Choo, Colin Boyd, and Yvonne Hitchcock.
\newblock Errors in computational complexity proofs for protocols.
\newblock In Bimal~K. Roy, editor, {\em ASIACRYPT}, volume 3788 of {\em Lecture
  Notes in Computer Science}, pages 624--643. Springer, 2005.

\bibitem{DentA:another}
Alexander~W Dent.
\newblock Fundamental problems in provable security and cryptography.
\newblock {\em Philosophical Transactions of the Royal Society A: Mathematical,
  Physical and Engineering Sciences}, 364(1849):3215--3230, 2006.

\bibitem{DH}
Whitfield Diffie and Martin~E. Hellman.
\newblock New directions in cryptography.
\newblock {\em IEEE Transactions on Information Theory}, IT-22(6):644--654,
  1976.

\bibitem{Dolev-Even-Karp}
Danny Dolev, Shimon Even, and Richard~M. Karp.
\newblock On the security of ping-pong protocols.
\newblock In {\em CRYPTO}, pages 177--186, 1982.

\bibitem{Dolev-Yao}
Danny Dolev and Andrew~C. Yao.
\newblock On the security of public key protocols.
\newblock {\em IEEE Transactions on Information Theory}, 29(2):198--208, 1983.

\bibitem{PavlovicD:Qabs12}
{Dusko Pavlovic}.
\newblock Geometry of abstraction in quantum computation.
\newblock {\em Proceedings of Symposia in Applied Mathematics}, 71:233--267,
  2012.
\newblock arxiv.org:1006.1010.

\bibitem{Freyd:abelian}
Peter Freyd.
\newblock {\em Abelian Categories: an Introduction to the Theory of Functors}.
\newblock Harper and Row, 1964.

\bibitem{Goldreich:bookr}
Oded Goldreich.
\newblock {\em Foundations of Cryptography}.
\newblock Cambridge University Press, 2000.

\bibitem{Goldwasser-Micali82}
Shafi Goldwasser and Silvio Micali.
\newblock Probabilistic encryption \& how to play mental poker keeping secret
  all partial information.
\newblock In {\em STOC '82: Proceedings of the fourteenth annual ACM symposium
  on Theory of computing}, pages 365--377, New York, NY, USA, 1982. ACM Press.

\bibitem{Goldwasser-Micali84}
Shafi Goldwasser and Silvio Micali.
\newblock Probabilistic encryption.
\newblock {\em J. Comput. Syst. Sci.}, 28(2):270--299, 1984.

\bibitem{Grillet:semigroups}
Pierre~Antoine Grillet.
\newblock {\em Semigroups: an introduction to the structure theory}.
\newblock Marcel Dekker, Inc., 1995.

\bibitem{Joyal-Street:geometry}
Andr\'{e} Joyal and Ross Street.
\newblock The geometry of tensor calculus {I}.
\newblock {\em Adv. in Math.}, 88:55--113, 1991.

\bibitem{Katz-Lindell:book}
Jonathan Katz and Yehuda Lindell.
\newblock {\em Introduction to Modern Cryptography}.
\newblock Chapman \& Hall/CRC Series in Cryptography and Network Security.
  Chapman \& Hall/CRC, 2007.

\bibitem{KellyGM:clubs}
Gregory~Max Kelly.
\newblock On clubs and doctrines.
\newblock In Gregory~Max Kelly, editor, {\em Category Seminar. Sydney 1972/73},
  pages 181--256. Springer-Verlag, Berlin, 1974.

\bibitem{another-two}
Neal Koblitz and Alfred Menezes.
\newblock Another look at {"Provable Security". II}.
\newblock In Rana Barua and Tanja Lange, editors, {\em INDOCRYPT}, volume 4329
  of {\em Lecture Notes in Computer Science}, pages 148--175. Springer, 2006.

\bibitem{another}
Neal Koblitz and Alfred Menezes.
\newblock Another look at {"Provable Security"}.
\newblock {\em J. Cryptology}, 20(1):3--37, 2007.

\bibitem{another-brave}
Neal Koblitz and Alfred Menezes.
\newblock The brave new world of bodacious assumptions in cryptography.
\newblock {\em Notices of the American Mathematical Society}, 57(3):357--365,
  March 2010.

\bibitem{LambekJ:abacus}
Joachim Lambek.
\newblock How to program an infinite abacus.
\newblock {\em Canad. Math. Bull.}, 4(3):295--302, 1961.

\bibitem{LambekJ:rings}
Joachim Lambek.
\newblock {\em Lectures on Rings and Modules}.
\newblock Blaisdell Publishing Co., 1966.

\bibitem{LambekJ:Advances}
Joachim Lambek.
\newblock From types to sets.
\newblock {\em Adv. in Math.}, 36:113--164, 1980.

\bibitem{Lambek-Scott:book}
Joachim Lambek and Philip~J. Scott.
\newblock {\em Introduction to higher order categorical logic}, volume~7 of
  {\em Cambridge Stud. Adv. Math.}
\newblock Cambridge University Press, New York, NY, USA, 1986.

\bibitem{MacLane:Homology}
Saunders Mac~Lane.
\newblock {\em Homology}.
\newblock Springer-Verlag, 1963.

\bibitem{PavlovicD:mapsII}
Dusko Pavlovic.
\newblock Maps {II}: Chasing diagrams in categorical proof theory.
\newblock {\em J. of the IGPL}, 4(2):1--36, 1996.

\bibitem{PavlovicD:MSCS97}
Dusko Pavlovic.
\newblock Categorical logic of names and abstraction in action calculus.
\newblock {\em Math. Structures in Comp. Sci.}, 7:619--637, 1997.

\bibitem{PavlovicD:IC12}
Dusko Pavlovic.
\newblock Monoidal computer {I}: {Basic computability by string diagrams}.
\newblock {\em Information and Computation}, 2013.
\newblock to appear; arxiv:1208.5205.

\bibitem{PavlovicD:MFPS10}
Dusko Pavlovic and Catherine Meadows.
\newblock Bayesian authentication: {Quantifying} security of the {Hancke-Kuhn}
  protocol.
\newblock {\em E. Notes in Theor. Comp. Sci.}, 265:97 -- 122, 2010.

\bibitem{PavlovicD:CRN1}
Dusko Pavlovic and Vaughan Pratt.
\newblock The continuum as a final coalgebra.
\newblock {\em Theor. Comp. Sci.}, 280(1--2):105--122, 2002.

\bibitem{Rackoff-Simon91}
Charles Rackoff and Daniel~R. Simon.
\newblock Non-interactive zero-knowledge proof of knowledge and chosen
  ciphertext attack.
\newblock In Joan Feigenbaum, editor, {\em CRYPTO}, volume 576 of {\em Lecture
  Notes in Computer Science}, pages 433--444. Springer, 1991.

\bibitem{Shannon:Secrecy}
Claude~E. Shannon.
\newblock Communication theory of secrecy systems.
\newblock {\em Bell Systems Technical Journal}, 28:656--715, 1949.

\bibitem{ShoupV:OAEP}
Victor Shoup.
\newblock {OAEP} reconsidered.
\newblock In {\em Proceedings of the 21st Annual International Cryptology
  Conference on Advances in Cryptology}, CRYPTO '01, pages 239--259, London,
  UK, 2001. Springer-Verlag.

\end{thebibliography}
\bibliographystyle{plain}

\end{document}